\documentclass{jfm}

\usepackage{lineno}
\usepackage{graphicx}
\usepackage{epstopdf,epsfig}
\usepackage{newtxtext}
\usepackage{newtxmath}
\usepackage{natbib}
\usepackage{hyperref}
\hypersetup{
    colorlinks = true,
    urlcolor   = blue,
    citecolor  = blue,
}

\newcommand{\RomanNumeralCaps}[1]

\shorttitle{The droplet size and effective viscosity asymmetries in turbulent emulsions}
\shortauthor{L. Yi and others}

\title{Physical mechanisms for droplet size and effective viscosity asymmetries in turbulent emulsions}

\author{
	Lei Yi,\aff{1}
	Cheng Wang,\aff{1}
	Thomas van Vuren,\aff{2}
	Detlef Lohse,\aff{3,4}
	Fr\'ed\'eric Risso,\aff{5}
	Federico Toschi,\aff{2,6}
	\and 
	Chao Sun\aff{1,7}
	\corresp{\email{chaosun@tsinghua.edu.cn}}
}

\affiliation{
	\aff{1}Center for Combustion Energy, Key Laboratory for Thermal Science and Power Engineering of
Ministry of Education, Department of Energy and Power Engineering, Tsinghua University,
100084 Beijing, China	
	
	\aff{2}Department of Physics and Department of Mathematics and
	Computer Science, Eindhoven University of Technology, 5600 MB
	Eindhoven, The Netherlands
	
	\aff{3}Physics of Fluids Group and Max Planck Center Twente, MESA+ Institute and J.M. Burgers Center for Fluid Dynamics, University of Twente, P.O. Box 217, 7500AE Enschede, The Netherlands
	
	\aff{4}Max Planck Institute for Dynamics and Self-Organization, 37077 G\"ottingen, Germany
	
	\aff{5}Institut de M\'ecanique des Fluides de Toulouse (IMFT), Universit\'e de Toulouse, CNRS, Toulouse, France
	
	\aff{6}CNR-IAC, Via dei Taurini 19, 00185 Roma, Italy
	
	\aff{7}Department of Engineering Mechanics,
	School of Aerospace Engineering, Tsinghua University, Beijing
	100084, China
}

\date{\today}

\begin{document}
\maketitle

\begin{abstract}
By varying the oil volume fraction, the microscopic droplet size and the macroscopic rheology of emulsions are investigated in a Taylor-Couette (TC) turbulent shear flow. 
Although here oil and water in the emulsions have almost the same physical properties (density and viscosity), unexpectedly, we find that oil-in-water (O/W) and water-in-oil (W/O) emulsions have very distinct hydrodynamic behaviors, i.e., the system is clearly asymmetric.
By looking at the micro-scales, the average droplet diameter hardly changes with the oil volume fraction neither for O/W nor for W/O. However, for W/O it is about $50\%$ larger than that of O/W. At the macro-scales, the effective viscosity of O/W is higher when compared to that of W/O.
These asymmetric behaviors can be traced back to the presence of surface-active contaminants in the system. By introducing an oil-soluble surfactant at high concentration, remarkably, we recover the symmetry (droplet size and effective viscosity) between O/W and W/O emulsions. Based on this, we suggest a possible mechanism responsible for the initial asymmetry.
Next, we discuss what sets the droplet size in turbulent emulsions.
We uncover a $-6/5$ scaling dependence of the droplet size on the Reynolds number of the flow. 
Combining the scaling dependence and the droplet Weber number, we conclude that the droplet fragmentation, which determines the droplet size, occurs within the boundary layer and is controlled by the dynamic pressure caused by the gradient of the mean flow, as proposed by~\cite{levich1962physicochemical}, instead of the dynamic pressure due to turbulent fluctuations, as proposed by~\cite{kolmogorov1949breakage}.
The present findings provide an understanding of both the microscopic droplet formation and the macroscopic rheological behaviors in dynamic emulsification, and connects them.
\end{abstract}

\section{Introduction}
Emulsions, such as mixtures of oil and water, have numerous industrial applications, including enhanced oil recovery, liquid-liquid extraction, drug delivery systems, and food processing~\citep{mandal2010characterization,mcclements2007critical,maffi2021mechanisms}. 
We can distinguish two types of emulsions: oil droplets in water and water droplets in oil, which we abbreviate with O/W and W/O, respectively~\citep{salager2000current}. What emulsion type is realized depends on a number of variables, among which the dispersed phase volume fraction is determinant~\citep{zambrano2003emulsion}. 
Typically, by increasing the dispersed phase volume fraction, $\phi_{d}$, a point is reached where the system experiences a so-called catastrophic phase inversion, by which the dispersed phase suddenly becomes the continuous one and vice versa~\citep{piela2009phenomenological}.
The evolution from O/W to W/O (or vice versa) can be accompanied by a dramatic change of the emulsion properties, including its morphology, rheology, and stability~\citep{perazzo2015phase}. 
Various studies show that asymmetric behaviors between O/W and W/O emulsions can be found for both the phase inversion characteristics and hydrodynamic behaviors, such as the critical volume fraction for the phase inversion~\citep{pacek1994structure}, even when the densities and the viscosities of the two phases in an oil-water system are identical~\citep{kumar1996phase}. W/O emulsions in a gravity settler were found to separate much more rapidly than their O/W counterparts~\citep{kato1991types}. 
The same holds for emulsion in a Taylor-Couette turbulent flow~\citep{bakhuis2021catastrophic}.
Additionally, it is found that O/W and W/O emulsions have different structures for a volume fraction $\phi_{d}>25\%$ of the dispersed phase~\citep{pacek1994structure}.
These experimental findings of the asymmetry in emulsions cannot be easily explained within the scope of existing models~\citep{kumar1996phase}.
Although some models (e.g., charged droplet model) have been proposed to account for the above mentioned observations~\citep{tobin1999modeling,kumar1996phase}, the understanding of asymmetric behaviors between O/W and W/O emulsions is still very limited.

Turbulent emulsions are complex physical systems, characterized by a dynamical coupling between small-scale droplets, large-scale flow and rheology. 
In the low-volume-fraction regime, droplet fragmentation is generally caused by the turbulent stress while the presence of droplets hardly affects the continuous phase~\citep{afshar2013liquid}.
The study of the droplet size in a turbulent flow can be traced back to~\cite{kolmogorov1949breakage} and~\cite{hinze1955fundamentals}, who attributed the droplet break-up to turbulent fluctuations. Although the Kolmogorov-Hinze (K-H) theory has been validated in a variety of experimental and numerical studies on droplets or bubbles in a turbulent flow~\citep{risso1998oscillations,perlekar2012droplet,eskin2017modeling,rosti2019droplets}, it was found to have limitations, for example, in non-homogeneous turbulent flows~\citep{hinze1955fundamentals}.
In the high-volume-fraction regime (before phase inversion), the microscopic droplet structure (droplet size and distribution), generated by the turbulent stresses, has a strong feedback on the macroscopic properties (viscosity) of the turbulent emulsion~\citep{de2019effect,yi2021global}. It has been found that the effective viscosity of the emulsion increases with increasing the volume fraction of the dispersed phase, which is similar to what is found for the case of suspensions of solid particles~\citep{rosti2021shear,stickel2005fluid,guazzelli2018rheology}. However, when considering the statistics of deformation, coalescence, and breakup, the dynamics of the droplets in emulsions is expected to be much richer than that of solid particles in suspensions.

The problem becomes even more complicated when we consider turbulent emulsions in practical environmental and industrial applications, where the appearance of dirt and surfactant in liquids or on the interfaces has to be taken into account~\citep{soligo2020effect,bazazi2020retarding}.
The surfactant dynamics can strongly modify the evolution of a flowing emulsion. 
On the one hand, the surfactant directly changes the interfacial properties, affecting the interface deformation and collision rate~\citep{manikantan2020surfactant}. 
On the other hand, the presence of surfactant can alter the global properties of the emulsion, such as its rheology~\citep{kawaguchi2016silicone}. 
However, the current understanding of the physics of turbulent emulsions with surfactant addition is still limited.

In this work, we study the dynamics of the emulsion in a turbulent shear flow, with an oil volume fraction ranging from  $0\%$ to $100\%$. We focus on the dispersed droplet size as a microscopic observer and the effective viscosity of the emulsion as a macroscopic observer. By introducing a surfactant at a controlled concentration into the system, we aim to reveal the physical mechanism for the asymmetric behaviors between oil-in-water and water-in-oil emulsions. Furthermore, we uncover the breakup mechanism of the droplet in such turbulent emulsions, for which the droplet Weber number plays an crucial role.

\section{Experimental setup and procedure} 

\begin{figure}
	\centering
	\includegraphics[width = 0.5\textwidth]{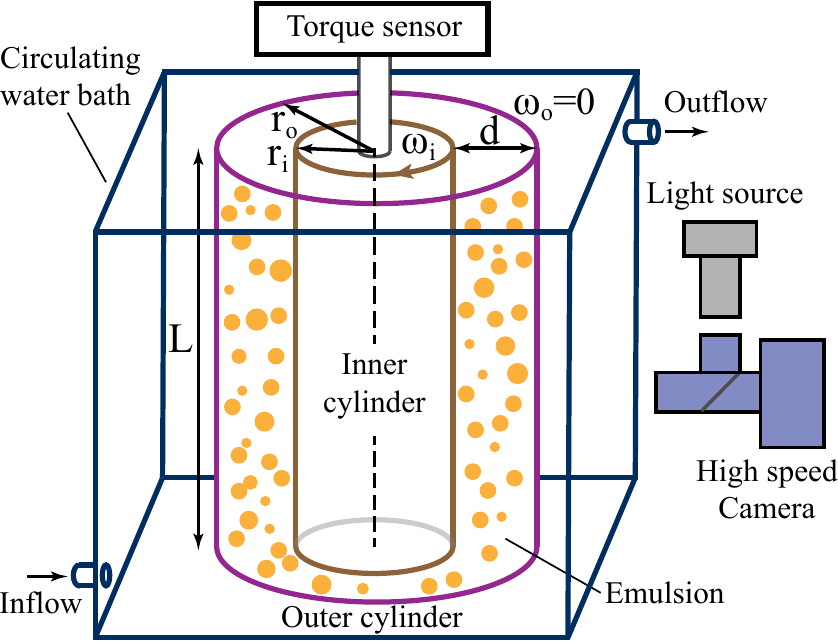}
	\caption{
		Sketch of the experimental set-up. The gap between the inner and the outer cylinders is filled with two immiscible liquids: silicone oil and ethanol-water. The emulsion is formed by rotating the inner cylinder at a given angular velocity $\omega_{i}$, while the outer cylinder is kept fixed. The torque of the inner cylinder is measured by a torque sensor. 
		A circulating water bath is used to maintain the working temperature at $\theta=22\pm0.1^\circ$C.
		A high-speed camera equipped with a long-distance microscope is used to capture the dispersed droplets in the flow.
	}
	\label{fig1} 
\end{figure}

In this study, the emulsion consists of two immiscible liquids: silicone oil (density $\rho_{o}=866\rm~kg/m^{3}$ and viscosity $\nu_{o}=2.1 \times10^{-6}\rm~m^{2}/s$) and an aqueous ethanol-water mixture ($\rho_{w}=860\rm~kg/m^{3}$, $\nu_{w}=2.4\times10^{-6}\rm~m^{2}/s$). 
The experiments were carried out in a Taylor-Couette (TC) system (see figure~\ref{fig1}(a)). 
The system has an inner cylinder radius $r_{i}=25\rm~mm$, an outer cylinder radius
$r_{o}=35\rm~mm$, and a gap $d=10\rm~mm$, giving a radius ratio of $\eta=r_{i}/r_{o}=0.71$. The height of the inner cylinder is $L=75\rm~mm$, so that the aspect ratio is $\Gamma=L/d=7.5$. The inner cylinder is made of aluminum, while the outer one is made of glass to enable optical measurements. 
Initially, the gap between the cylinders is filled by the ethanol-water mixture and the oil. Then, the inner cylinder is set in rotation at a constant angular velocity $\omega_{i}$, while the outer cylinder is kept fixed (i.e., $\omega_{o} = 0$). A strong turbulent shear flow is generated which generates an emulsion. After a certain time, the emulsion finally reaches a state where its statistical properties are steady.
Note that the density match of the two phases (oil and ethanol-water) eliminates the effect of the centrifugal force on the liquid distribution in the system (see appendix~\ref{appendix_A}).
A circulating water bath is used to maintain the working temperature at $\theta=22\pm0.1^\circ$C. The temperature gradient in the emulsion is negligible due to the efficient mixing induced by the turbulent fluctuations~\citep{vangils2011rsi,grossmann2016high}.
The control parameter of the Taylor-Couette flow is
the Reynolds number defined as: $Re = \omega_{i}r_id/\nu_{w}$, where $\omega_{i}$ is the imposed angular velocity of the inner cylinder, and $\nu_{w}$ is the viscosity of the ethanol-water. 
Here we also define a modified Reynolds number, $Re_{m} = \omega_{i}r_id/\nu_{eff}$, where $\nu_{eff}$ is the effective viscosity of the emulsion.
We measured the total torque exerted on the inner cylinder $T_{raw}$, which includes two parts: the torque contribution from the cylindrical sidewall surfaces (the TC flow), $T$, and the torque contribution from both the top and bottom end plates (the end flow), $T_{end}$. $T_{end}$ is measured using the same linearization method in previous  studies~\citep{hu2017significant,greidanus2011drag} (see appendix~\ref{appendix_B}). Thus, the torque contribution of the TC flow can be determined by $T = T_{raw} - T_{end}$.
Based on this, the dynamic response of the emulsion to the imposed rotation is characterized by the dimensionless torque: $G=T/2\pi L\rho\nu_{w}^{2}$, and a modified one: $G_{m}=T/2\pi L\rho\nu_{eff}^{2}$.

The dispersed oil (or ethanol-water) droplets in the emulsion were captured using the high-speed camera equipped with a long-distance microscope. Videos and images from experiments were analyzed for the drop size determination using ImageJ software and Matlab codes. The numerical average of the droplet diameter $\left<D\right>$ is used as the indicator of the droplet size in this study. To ensure achieving enough statistics, the average droplet diameter is calculated based on $\mathcal{O}(10^3)$ droplet samples.
Experiments were performed for various oil volume fractions, $\phi_{o}$, and Reynolds numbers, $Re$.

\begin{figure}
	\centering
	\includegraphics[width = 1\textwidth]{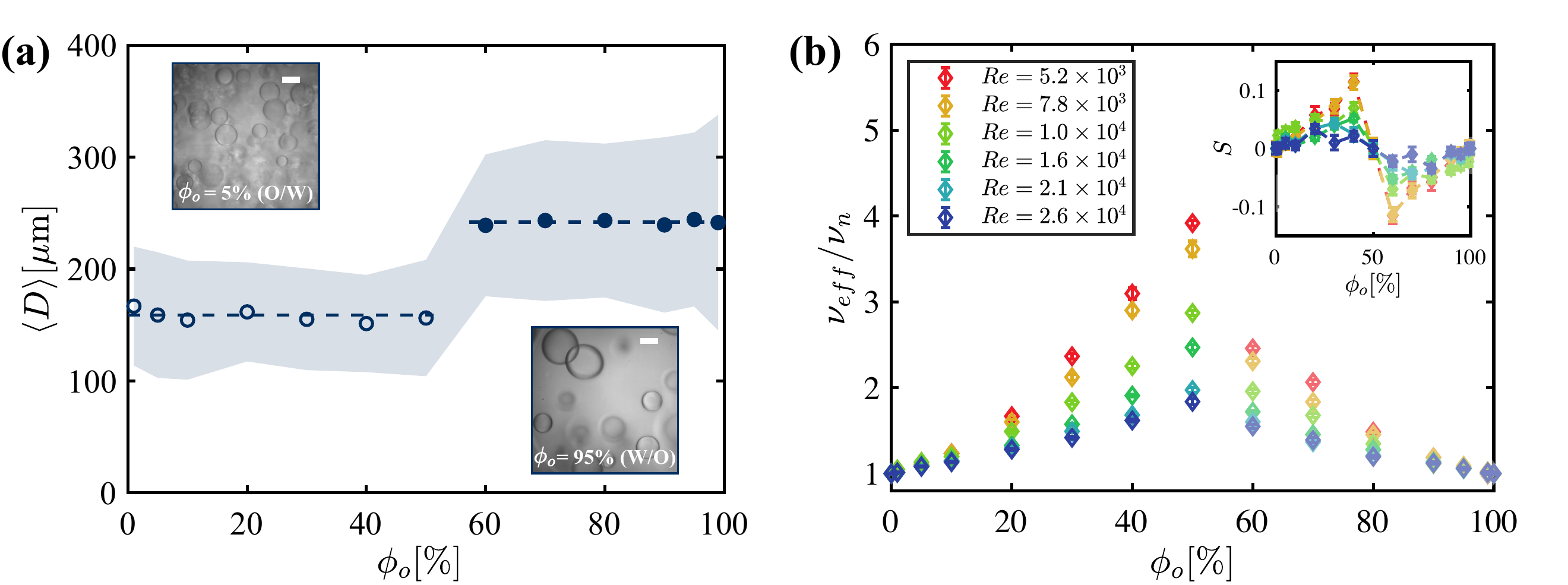}
	\caption{
		(a) Average droplet diameter as a function of the oil volume fraction. The droplet diameter hardly changes with the oil volume fraction for both O/W emulsions ($\phi_{o} \le 50\%$, hollow points) and W/O emulsions ($\phi_{o} \ge 60\%$, solid points). The ethanol-water droplets in W/O are found to be about $50\%$ larger than the oil droplets in O/W.
		The bluish band spans one standard deviation around the average diameter. The dashed lines denote the average values of the droplet diameters for O/W (or W/O).
		A typical O/W emulsion ($\phi_{o} = 5\%$) and a typical W/O emulsion ($\phi_{o} = 95\%$) are shown in the insets. The scalar bar represents $200 \rm\mu m$.
		(b) The normalized effective viscosity of the emulsion as a function of the oil volume fraction, $\phi_{o}$, for various Reynolds numbers, $Re$. The inset shows the asymmetry factor $S$ as a function of $\phi_{o}$ for various $\omega_{i}$. The asymmetric trend of the effective viscosity between O/W (left branch) and W/O (right branch) emulsions is more pronounced at low Reynolds numbers.
	}
	\label{fig2} 
\end{figure}

\section{Results and discussion}

\subsection{Asymmetric behaviors of the droplet size and the effective viscosity}
The size of the dispersed droplets in a turbulent emulsion characterizes the microscopic structure of the emulsion, which affects the macroscopic stability and rheology of the emulsion. 
Also, the volume fraction of the dispersed phase, $\phi_{d}$, determines both the micro-scale structure (droplet size) and, consequently, the macro-scale rheological behaviors (effective viscosity). 

Firstly, we focus on the effect of the oil volume fraction, $\phi_{o}$, on the droplet size, for a given Reynolds number of $Re = 5.2\times10^{3}$. 
The volume fraction of oil, $\phi_{o}$, is varied from 0 (ethanol-water mixture) to $100\%$ (pure oil) by fixing the volume of each phase initially put into the TC system. After emulsification, the final state of the emulsion is observed to be O/W for $\phi_{o} \le 50\%$ and W/O for $\phi_{o} \ge 60\%$. A phase inversion process, in which the continuous phase and the dispersed phase are exchanged~\citep{salager2000current}, is thus expected to occur in between these two volume fractions. In this range, the behavior of the system is too complex to allow for an accurate determination of the inversion point. In the following, we will only consider values of $\phi_{o}$ either below $50\%$ or larger than $60\%$, for which the nature of the dispersed and continuous phases are unambiguously determined.
In this study, we focus on the global and local properties of O/W and W/O emulsions in a turbulent flow.
Note that all emulsions we obtained are of a simple type, and we did not observe multiple emulsions, such as O/W/O or W/O/W~\citep{perazzo2015phase}.
Typical O/W emulsion for $\phi_{o} = 5\%$ and W/O emulsion for $\phi_{o} = 95\%$ are shown in inset images in figure~\ref{fig2}(a).
Under steady stirring conditions, the droplet size in the turbulent emulsion eventually shows a statistically stationary distribution, giving an average of the droplet diameter, $\left<D\right>$, as an indicator for the droplet size. 
The average droplet diameters for various oil fractions are shown in figure~\ref{fig2}(a).
For both O/W emulsions (left branch, $\phi_{o} \le 50\%$) and W/O (right branch, $\phi_{o} \ge 60\%$) emulsions, it is found that the droplet size is almost independent of the oil fraction, at fixed $Re$.
Remarkably, we find that the ethanol-water droplets in the right branch are about $50\%$ larger than the oil droplets in the left branch, indicating an obvious asymmetry of the droplet size between O/W and W/O emulsions.
One may think that this asymmetric behavior is due to the slight difference in physical properties of the two liquids used in experiments. However, we note that the densities of these two liquids are too close to account for the observed asymmetry. The interfacial tension between the two immiscible liquids is also identical for O/W and W/O emulsions.
What about the viscosity? The viscosity of the silicone oil, $\nu_{o}=2.1 \times10^{-6}\rm~m^{2}/s$, is slightly lower than that of the ethanol-water, $\nu_{w}=2.4 \times10^{-6}\rm~m^{2}/s$, at the experimental temperature of $\theta=22^\circ$C, while it has been found that at least an order of magnitude difference between the viscosities of the two phases could change the droplet size by a measurable amount~\citep{pacek1994structure}.
Moreover, additional experiments we performed show that the asymmetry of the droplet size remains, even when we eliminate viscosity difference by adjusting the working temperature (see appendix~\ref{appendix_D} for more details).
Thus, the small viscosity difference cannot account for the observed asymmetry of the droplet size in these experiments. The origin of the asymmetry of the droplet size must therefore have another origin. 

Apart from the droplet size, also the effective viscosity of the emulsion shows an asymmetric behavior. 
For various oil volume fractions ($0\% \le \phi_{o} \le 100\%$), we measured the effective viscosity of the emulsion, $\nu_{eff}$, which is calculated using a method that has been recently proposed for viscosity measurements in a turbulent Taylor-Couette flow~\citep{bakhuis2021catastrophic,yi2021global}. 
As well known, an effective power-law dependence $G \propto Re^{\alpha}$ holds in the single-phase Taylor–Couette turbulent flow, where the power-law exponent $\alpha$ depends on the regime of the Reynolds number~\citep{grossmann2016high}. Indeed, if the modified Reynolds number and dimensionless torque are used, the current two-phase emulsion flow still follows the effective power-law dependence: $G_{m} \propto Re_{m}^{\alpha}$, where $\alpha = 1.58$ in this study (see appendix~\ref{appendix_B}). Based on this relation, we can calculate the effective viscosity of the emulsion, $\nu_{eff}$.
The detailed calculation of the effective viscosity is documented in appendix~\ref{appendix_B}. The results of the effective viscosity are shown in figure~\ref{fig2}(b), which can also be divided into two parts: O/W emulsions for $\phi_{o} \le 50\%$ (left branch) and W/O emulsions $\phi_{o} \ge 60\%$ (right branch).
Here the effective viscosity is characterized by its normalized value, $\nu_{eff}/\nu_{n}$, where $\nu_{n} = \nu_{w}$ for O/W ($\phi_{o} \le 50\%$) and $\nu_{n} = \nu_{o}$ for W/O ($\phi_{o} \ge 60\%$).
For each branch, the effective viscosity increases with the increasing dispersed phase volume fraction $\phi_{d}$ for all Reynolds numbers. 
Note that the dispersed phase refers to oil for O/W or ethanol-water for W/O.
The effective viscosity has only a weak dependence on the dispersed phase volume fraction in the dilute regime (i.e., for $\phi_{d} < 5\%$), while it displays a stronger dependence at larger dispersed phase volume fractions. The increase of the effective viscosity with increasing $\phi_{d}$ originates from the hydrodynamic or contact interactions between dispersed droplets, as it is observed in similar turbulent droplet dispersions~\citep{pouplin2011wall} and in solid particle suspensions~\citep{guazzelli2018rheology}. 
Furthermore, the effective viscosity is found to decrease with the increasing Reynolds number for a given $\phi_{d}$, indicating that the turbulent emulsion somehow shows a shear shinning behavior~\citep{yi2021global}.
Though the qualitative trend of the effective viscosity versus the dispersed phase volume fraction is similar for both the left and the right branches, an asymmetry of the effective viscosity between O/W and W/O emulsions is measured. The effective viscosity of O/W (left branch) is found to be higher than that of W/O (right branch) for a given Reynolds number, particularly for the case of the high dispersed phase volume fraction (see figure~\ref{fig2}(b)).
To quantitatively represent the degree of asymmetry, we define an asymmetry factor as:

\begin{equation}
	S = \frac{\Big(\frac{\nu_{eff}}{\nu_{n}}\Big)_{\phi_{o}} - \Big(\frac{\nu_{eff}}{\nu_{n}}\Big)_{1-\phi_{o}}}{\Big(\frac{\nu_{eff}}{\nu_{n}}\Big)_{\phi_{o}} + \Big(\frac{\nu_{eff}}{\nu_{n}}\Big)_{1-\phi_{o}}},
\end{equation}
where the subscripts $\phi_{o}$ and $1 - \phi_{o}$ denote the emulsion at $\phi_{o}$ and at $1 - \phi_{o}$, respectively. An asymmetry factor $S$ deviating from $0$ indicates asymmetry. The asymmetry factor, as a function of the oil volume fraction, is shown as the inset of figure~\ref{fig2}(b). It is found that the asymmetry decreases with the increasing Reynolds number. In addition, the asymmetric trend between O/W and W/O is more pronounced for high dispersed phase volume fractions.
Since it was already found that the droplet size has a dramatic influence on the emulsion rheology~\citep{pal1996effect}, the macroscopic asymmetry of the effective viscosity between O/W and W/O could be connected to the microscopic asymmetric behavior of the droplet size.

\subsection{Recovering the asymmetry between O/W and W/O emulsions using surfactant}

\begin{figure}
	\centering
	\includegraphics[width = 1\textwidth]{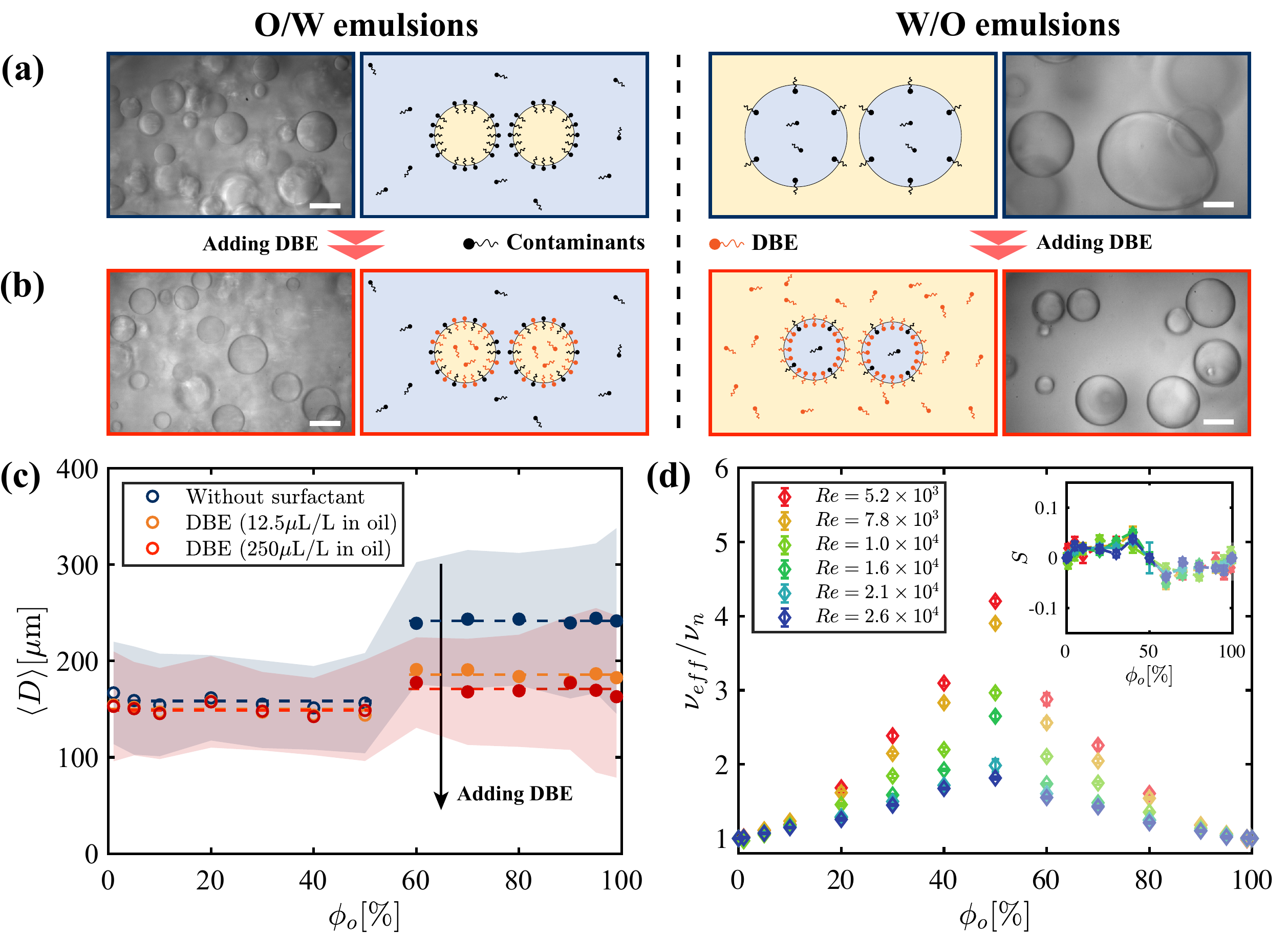}
	\caption{
		(a-b) Schematic diagrams and microscopic images of the effects of surfactants (or surface-active contaminants) on the emulsions. 
		Part (a) depicts the emulsions without adding surfactant. The asymmetric behaviors of the droplet size between O/W emulsions and W/O emulsions are observed due to the difference in the concentration of contaminants on the interfaces. The deformation of the droplet is observed clearly in W/O emulsions. 
		Part (b) represents emulsions with adding DBE at $250~\mu \text{L/L}$. The surfactant is adsorbed into the interface at saturation for both O/W and W/O emulsions. The symmetry of droplet size is almost recovered.
		Snapshots of the emulsion at an oil volume fraction of $\phi_{o}=5\%$ are used to represent O/W emulsions, while W/O emulsions are denoted by snapshots of the emulsion at $\phi_{o}=95\%$. For all snapshots here, $Re = 5.2\times10^{3}$. The scalar bar represents $200~\rm\mu m$.
		(c) The droplets size in emulsions for various oil volume fractions, with and without adding the surfactant, for $Re = 5.2\times10^{3}$. The droplet size almost does not change with the oil volume fraction for both O/W ($\phi_{o} \le 50\%$) and W/O ($\phi_{o} \ge 60\%$) emulsions. The asymmetric trends of the droplet size between O/W and W/O are found to disappear when adding surfactant DBE into the emulsion systems. Note that $12.5~\mu \text{L/L}$ and $250~\mu \text{L/L}$ here refer to the concentrations of DBE in oil.
		The dashed lines represent the average values of the droplet diameters for O/W (or W/O).
		The colored bands span the average diameter ± one standard deviation. For the sake of clarity, the colored band for the case using $12.5~\mu \text{L/L}$ DBE is not present here.
		(d) The normalized effective viscosity of the emulsion for the oil volume fraction from $\phi_{o}=0\%$ to $\phi_{o}=100\%$, with adding surfactant DBE at $250~\mu \text{L/L}$, at various Reynolds numbers. The inset shows the asymmetry factor $S$ as a function of the oil volume fraction. The asymmetric behavior of the effective viscosity between O/W emulsions and W/O emulsions becomes considerably smaller due to the presence of surfactant DBE at saturation.
	}
	\label{fig3} 
\end{figure}

We hypothesize that the possible reason for the asymmetry of the droplet size between O/W and W/O emulsions is the presence of surface-active contaminants.
These surface-active contaminants are widely found on the liquid-liquid interface in practical environments, which is focused on by various studies related to interfacial phenomena~\citep{calvo2019coalescence,de2015coalescence,de2001some}. On the one hand, these surface-active contaminants can modify the interfacial properties, yielding the change of the droplet size in the emulsion~\citep{bazazi2020retarding,manikantan2020surfactant}. On the other hand, the solubility of these contaminants is usually different in the oil phase and the aqueous phase~\citep{kawaguchi2016silicone}. The preferential solubility can induce a different distribution of contaminants and different interfacial properties in O/W and W/O, which could be the source of the asymmetric behaviors.

To investigate the effect of surfactants on the asymmetric behavior of turbulent emulsions, an effective way is to add a controlled amount of a selected surfactant into the system. 
Firstly, we perform experiments using a kind of oil-soluble nonionic surfactant: dimethylsiloxane block copolymer (30-35\% Ethylene Oxide). For convenience, we use its abbreviation (DBE) from the manufacturer. 
For the purpose of the present study, two contrasted concentrations of DBE in oil are selected. One is $12.5~\mu \text{L/L}$, which is comparable to the critical micelle concentration (CMC) of DBE in water (around $13~\mu \text{L/L}$)~\citep{rheingans2000nanoparticles}, and the second, 20 times larger, is $250~\mu \text{L/L}$. 
The DBE is well mixed with the oil before each experiment.
Two microscopic images of O/W and W/O emulsions with adding $250~\mu \text{L/L}$ DBE are shown in figure~\ref{fig3}(b). Figure~\ref{fig3}(c) shows the results of the droplet size in emulsions using DBE, for various oil volume fractions. Here, the Reynolds number is fixed at $Re = 5.2\times10^{3}$.
The measured droplet sizes for emulsions using $12.5~\mu \text{L/L}$ and $250~\mu \text{L/L}$ DBE are denoted by yellow marks and red marks in figure~\ref{fig3}(c), respectively.
It is found that DBE only slightly reduces the droplet size in O/W emulsions (left branch) when compared to what has been found in emulsions without adding surfactant (blue marks). 
But for W/O emulsions (right branch), the droplet size decreases with the increasing concentration of DBE. When $250~\mu \text{L/L}$ DBE is used, remarkably, the droplet size difference between O/W and W/O emulsions is eliminated (see red marks in figure~\ref{fig3}(c)). Consequently, we nearly recover the symmetry of the droplet size between O/W and W/O by adding oil-soluble surfactant DBE at high concentration into the emulsion system.

The above results can be explained using the schematic diagrams in figure~\ref{fig3}(a-b). 
We first focus on the cases without adding surfactant.
In the practical environment, even without adding any surfactant, the emulsion inevitably contains some surface-active contaminants~\citep{soligo2020effect,bazazi2020retarding} (black indicators in figure~\ref{fig3}(a)), which measurably originate from the wall of the container in this study. These surface-active contaminants are measurably preferentially soluble in aqueous ethanol-water and act as a surfactant. 
As illustrated in the left part of figure~\ref{fig3}(a) (O/W), the surface-active contaminants from the wall dissolve into the continuous phase of ethanol-water.
These contaminants are then adsorbed into the liquid-liquid interface and modify the oil droplet size for two reasons.
On the one hand, contaminants on the surface suppress droplet coalescence, which is known to be a common effect of surfactants on emulsion systems~\citep{baret2012surfactants,dai2008mechanism,ha2003effect,cristini1998near}.
On the other hand, these surface-active contaminants fully cover the surface of oil droplets, inducing a reduction of the interfacial tension~\citep{manikantan2020surfactant}. Thus, the breakup of droplets could be promoted.
The effect of contaminants on the breakup and coalescence of the droplets finally reflects on the smaller size of oil droplets in O/W.
As the interfacial tension is found to only slightly decrease with the concentration of DBE (see appendix~\ref{appendix_A}), it is reasonable to assume that the inhibition of the droplet coalescence is the dominant factor affecting the droplet size here.
However, when oil is the continuous phase (W/O) as shown in the right part of figure~\ref{fig3}(a), ethanol-water droplets embedded within the oil are not in contact with walls. Therefore, only a few surface-active contaminants are adsorbed into the interface. The cleaner liquid-liquid interface brings less inhibition to the coalescence, yielding the larger droplet size for the given turbulent strength (see figure~\ref{fig3}(a)). 

The effective viscosity difference between O/W and W/O could be related to the deformability of the dispersed phase, which is closely connected to the droplet size~\citep{pal1996effect,van2013importance,verschoof2016bubble,saiki2007effects}.
When compared to the larger ethanol-water droplets, the small and non-deformable oil droplets could yield a larger resistance to the straining component of the shearing flow, thereby increasing the effective viscosity of the emulsion~\citep{stickel2005fluid,guazzelli2018rheology,bakhuis2018finite}. 
In addition, it was already found that surface-active contaminants give rise to surface-excess rheology as compared to a clean surface~\citep{brenner2013interfacial,elfring2016surface,langevin2014rheology}, which is also called interfacial rigidification effect~\citep{erni2011deformation}. 
Thus, oil droplets covered by contaminants are less deformable, which brings an extra contribution to the effective viscosity of O/W~\citep{guazzelli2018rheology}. 
Furthermore, surface-active contaminants could modify the dynamics of the emulsion through the hydrodynamic coupling interaction between the oil-droplet surface and the surrounding flow~\citep{baret2012surfactants}.
As a result, the surface-active contaminants account for the measured higher effective viscosity of O/W (left branch) than that of W/O (right branch) (see figure~\ref{fig3}(b)).

Next, we consider the experimental results using the oil-soluble surfactant DBE. For the case of O/W emulsion shown in the left part of figure~\ref{fig3}(b), some DBE added in the system is competitively adsorbed into the interface. As the interface is already saturated due to the contaminants, the interfacial properties show no significant change when adding DBE, which is consistent with the previous result that the droplet sizes for O/W emulsions only slightly decrease when DBE is added. 
When the oil is the continuous phase (W/O) as shown in the right part of figure~\ref{fig3}(b), the surface of ethanol-water droplets is expected to be mostly covered by DBE adsorbed from oil, i.e, the surface is at saturation ($250~\mu \text{L/L}$ case). Therefore, the coalescence of droplets is now inhibited. 
Consequently, the asymmetric trend of the droplet size between O/W and W/O is eliminated using DBE (figure~\ref{fig3}(c)). 

In this part, we focus on the effect of surfactant on another feature of the emulsion system: the effective viscosity.
As shown in figure~\ref{fig3}(d), it is found that the symmetry of the effective viscosity between O/W (left branch) and W/O (right branch) emulsions is partially recovered using $250~\mu \text{L/L}$ DBE. 
This is clearly indicated by the asymmetry factor $S$ close to $0$ (see the inset of figure~\ref{fig3}(d)), which is expected to be mainly attributed to the recovery of the symmetry of the droplet size using DBE (see figure~\ref{fig3}(c)). 
It should be noted that the symmetry of the effective viscosity is not fully recovered. In general, there are always some differences between O/W and W/O emulsions, such as the distribution of the surfactant in the flow. 
The effective viscosity for each case is found to be slightly larger than that for its corresponding case without adding surfactant (compare figure~\ref{fig3}(d) to figure~\ref{fig3}(b)). The reason could be that the polymeric surfactant (DBE in this study) with high molecular weight enhances the interfacial rigidification effect of the droplet surface when compared to the case with only contaminants~\citep{erni2011deformation}. Indeed, the copolymer molecules of surfactant could form shell-like structures around the drops~\citep{sundararaj1995drop}. This increases the resistance of the droplet to the surrounding flow, yielding the extra contribution of the viscous dissipation of the flow. 

\subsection{The dependence of the droplet size on the Reynolds number}

\begin{figure}
	\centering
	\includegraphics[width = 1\textwidth]{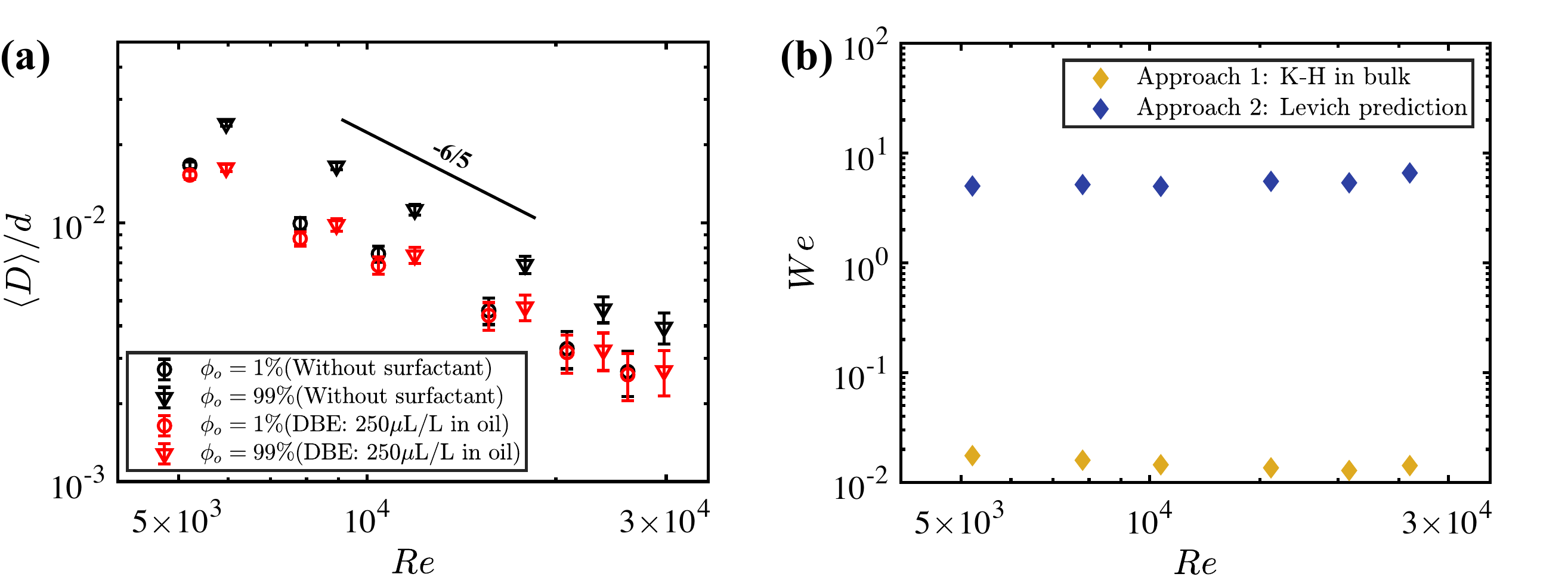}
	\caption{
		(a) The average droplet diameter normalized by the gap width as a function of the Reynolds number for oil volume fractions of $\phi_{o}=1\%$ and of $99\%$. The black and red symbols denote droplet sizes in emulsions using no surfactant and using $250~\mu \text{L/L}$ DBE, respectively. The error bars are based on the errors of the edge detection.
		The black lines denote the scaling prediction of the Kolmogorov-Hinze theory, $\left<D\right>/d \sim Re^{-6/5}$. The asymmetry of the droplet size in O/W ($\phi_{o}=1\%$) and W/O ($\phi_{o}=99\%$) emulsions is removed using DBE at saturation.
		(b) The Weber number as a function of the Reynolds number in a log-log plot for the two theoretical approaches. The yellow and blue diamonds denote the results using K-H theory based on the energy dissipation rate in the bulk area and that using the Levich's theory, respectively. O/W emulsions at $\phi_{o}=1\%$ without surfactant are considered here.
	}
	\label{fig4} 
\end{figure}
As the droplet size and the associated asymmetry have been discussed for various oil fractions, the next question is what sets the droplet size in turbulent emulsions.
In this part, we study the dependence of the droplet size on the Reynolds number, at the low volume fraction of the dispersed phase (i.e., $\phi_{o}=1\%$ and $\phi_{o}=99\%$). 
Since the volume fraction of the dispersed phase is very low in these cases, the viscosity of the emulsion is approximately equal to that of the continuous phase, giving $Re_{m} \approx Re$ and $G_{m} \approx G$.
We firstly consider the cases at $\phi_{o}=1\%$ without using surfactant.
Note that the coalescence of droplets is inhibited due to the surface-active contaminants on the droplet surface. Therefore, the droplet size is mainly determined by the turbulent breakup mechanism. 
As shown in figure~\ref{fig4}(a) by the black circles, the droplet size normalized by the gap, $\left<D\right>/d$, is found to have a scaling dependence on the Reynolds number, $Re$, with an effective exponent of $-1.18\pm0.05$ obtained by a direct fit for $\phi_{o}=1\%$. 

We now explore the physical mechanism behind the scaling dependence of the droplet size on the Reynolds number.
According to the Kolmogorov-Hinze theory, the droplet formation in a turbulent flow is determined by the competition between the deforming external dynamic pressure force (turbulent fluctuations) and the resisting interfacial tension over the droplet surface~\citep{kolmogorov1949breakage,hinze1955fundamentals}, of which the ratio is usually indicated by the droplet turbulent Weber number: $We = \rho \overline{\delta u^2} D/\gamma$, where $\rho$ is the density of the continuous phase, $\overline{\delta u^2}$ is the mean-square velocity difference over a distance equal to the droplet diameter $D$, and $\gamma$ is the interfacial tension between the two phases~\citep{risso1998oscillations}.
If the droplet diameter $D$ belongs to the inertial turbulent sub-range, $\overline{\delta u^2}$ could be expressed as a function of the local energy dissipation rate: $\overline{\delta u^2} = C_{1}(\varepsilon D)^{2/3}$, where the constant $C_{1}\approx 2$ according to Batchelor~\citep{batchelor1953theory}. This yields the Weber number as: $We = 2\rho\varepsilon^{2/3}D^{5/3}/\gamma$.
The force balance implies the existence of a critical value of the Weber number beyond which breakup occurs~\citep{hinze1955fundamentals}, and this value is found to be of order of unity (i.e. $We\sim\mathcal{O}(1)$) in various studies~\citep{hesketh1991bubble,risso1998oscillations,lemenand2017turbulent}.
Thus, the prediction of the maximum stable droplet size in a homogeneous and isotropic turbulent flow can be given by: $D_{max} = C(\rho/\gamma)^{-3/5}\varepsilon^{-2/5}$ ($C$ is a constant coefficient), which is the main result of the work by~\cite{hinze1955fundamentals}.
Moreover, various studies have shown that the average droplet diameter, $\left<D\right>$, can be used as the indicator of the droplet size in the Kolmogorov-Hinze  prediction~\citep{boxall2012droplet,perlekar2012droplet,lemenand2003droplets}.

Firstly, we speculate that the droplet size could be dominated by the turbulent fluctuations in the bulk flow of the system, where most droplets distribute. The local energy dissipation rate in the bulk can be estimated as $\varepsilon_b \sim u_{_T}^{3}/\ell$, where $u_{_T}$ and $\ell$ are the typical velocity fluctuation and the characteristic length scale of the flow~\citep{ezeta2018turbulence}. As the typical velocity fluctuation can be expressed as $u_{_T} \sim \omega_{i}r_{i} \sim Re\cdot \nu /d$~\citep{van2012optimal}, we obtain $\varepsilon_b \sim Re^3\nu^3/d^4$ by assuming $\ell \sim d$.
Inserting $\varepsilon_b$ into the K-H prediction, the scaling dependence of the droplet size on the Reynolds number is obtained as $\left<D\right>/d \sim Re^{-6/5}$, which agrees well with the experimental results for $\phi_{o} = 1\%$ without surfactant (see black circles in figure~\ref{fig4}(a)). 
However, the discussion above is only a simple analysis based on the scaling law. 
A further quantitative study on the droplet formation in a turbulent flow needs to consider the Weber number, which can be calculated as $We = 2\rho_{w}\varepsilon_b^{2/3}D^{5/3}/\gamma$, where $\varepsilon_b$ can be estimated as $\varepsilon_b \approx 0.1T\omega_{i}/[\pi(r_{o}^{2}-r_{i}^{2})L\rho_{w}]$ in a TC turbulent flow~\citep{ezeta2018turbulence}. As shown in figure~\ref{fig4}(b) (yellow diamonds, approach 1), the Weber number ranges between $0.013$ to $0.018$, two orders of magnitude smaller than the critical value obtained in previous studies, suggesting that the bulk of the system, where most droplets flow around, is not the place where the droplet size is determined.

Indeed, the droplet breakup is most often observed close to the area where the most intense stress participates in the deformation~\citep{hesketh1991experimental,afshar2013liquid}. 
Considering that the coalescence is almost inhibited in the current system, the droplet size is mainly dominated by the place where small droplets are generated.
Thus, the droplet size is expected to be dominated by the boundary layer area close to the wall, where the K-H theory has some limitations. 
The resulting droplet Weber number is too small as well when using the local energy dissipation rate near the wall, suggesting that the K-H theory is not appropriate for modeling the droplet size in the present system (see appendix~\ref{appendix_C} for details).

A prediction of the droplet size in the non-homogeneous turbulent flow past a solid wall was proposed by~\cite{levich1962physicochemical}, who gave the dynamic pressure force exerted on the two sides of the droplet using the logarithmic distribution of the mean velocity in the boundary layer. Note that the Reynolds number in the current study is in the interval where the logarithmic mean velocity distribution in the boundary layer can exist~\citep{huisman2013logarithmic}. According to~\cite{levich1962physicochemical}, the droplet diameter can be written as $\left<D\right> = 2\sqrt{\gamma \nu /(25\rho_{w} u_{\ast}^{3})}$, where we use the shear velocity: $u_{\ast} = \sqrt{\tau_{w}/\rho_{w}} = \sqrt{T/(2\pi\rho r_{i}^{2}L)}$. 
Using the effective scaling of $G \sim Re^{1.58}$ obtained in the current system, the scaling dependence of the droplet diameter on the Reynolds number is derived as $\left<D\right>/d \sim Re^{-1.19}$, where the exponent $-1.19$ is very close to the $-6/5$ in the K-H prediction and agrees again with the scaling dependence observed in experiments for $\phi_{o} = 1\%$. 
Based on the Levich theory, we also calculate the Weber number as the ratio of the dynamic pressure force induced by the mean flow to the interfacial tension: $We = 25\rho_{w} u_{\ast}^{3}\left<D\right>^{2}/(2\nu_{w}\gamma)$. 
As shown in figure~\ref{fig4}(b) (blue diamonds, approach 2), the Weber number for the Levich prediction is about $5$, which is consistent with the critical value for the droplet breakup in a turbulent flow~\citep{risso1998oscillations,lemenand2017turbulent}. 
The comparison of the Weber numbers based on the energy dissipation rate and that based on Levich’s theory lead to the conclusion that the droplet fragmentation, which determines the droplet size, occurs within the boundary layer and is controlled by the dynamic pressure caused by the gradient of mean flow, in agreement with the mechanism originally proposed by Levich.
Note that this conclusion also requires that the boundary layer thickness is larger than the droplets diameter, which is supported by the fact that boundary layer thickness is estimated as $5$ times the droplet size in this study (see appendix~\ref{appendix_C}).

The discussion of the two approaches above is based on the droplet size at $\phi_{o} = 1\%$, whereas the droplet size at $\phi_{o}=99\%$ is found to follow the same $-6/5$ scaling dependence (see black triangles in figure~\ref{fig4}(a)), indicating the robustness of the scaling law. Furthermore, this figure also shows the existence of the asymmetry of the droplet size between O/W and W/O at high $Re$, at least for low dispersed phase volume fractions (i.e., $\phi_{o}=1\%$ and $\phi_{o}=99\%$). 
Considering that the droplet size for W/O is about $50\%$ larger than that for O/W, only a slight variation of the Weber number is expected at $\phi_{o} = 99\%$ as compared to the case at $\phi_{o} = 1\%$. Therefore, all the qualitative conclusions given above are valid for emulsions at $\phi_{o} = 99\%$ as well.
In this case, the droplet coalescence needs to be considered.
The fact that the breakup theory, without accounting for coalescence, describes well the experimental result is particularly interesting. Indeed, for a steady state can be finally reached, the coalescence rate of droplets has to be equal to the breakup rate of coalesced droplets. We therefore observed an average size that is larger than that predicted by the breakup theory, but which remains proportional to it. 
Note that the interpretation given here is still not complete, and the results remain open for discussion.
\textcolor{black}{Additionally, the finding that the droplet formation is controlled by the boundary layer also provides a reasonable explanation for the observations that the droplet size hardly depends on the dispersed phase volume fraction $\phi_{d}$ (see figure~\ref{fig4}(a)). The droplets that are generated close to the wall, where similar mean velocity gradient could distribute for various $\phi_{d}$, are expected to have similar size. 
Therefore, the similar droplet size is observed at various $\phi_{d}$ in the entire system for O/W (or W/O) emulsions.}

Next, we turn to the results using surfactant. Experiments using $250~\mu \text{L/L}$ DBE are performed, and the results are shown in figure~\ref{fig4}(a). We note that the scaling dependence of the droplet size on the Reynolds number remains, suggesting the robustness of the scaling law for turbulent emulsions containing DBE.
For the case of $\phi_{o}=1\%$, it is found that the droplet size only slightly decreases due to DBE. However, the droplet size at $\phi_{o}=99\%$ shows a dramatic reduction to a value close to that at $\phi_{o}=1\%$, yielding the elimination of the asymmetry of the droplet size, for various Reynolds numbers. 
Since we have found that the recovery of the symmetry using DBE is due to the inhibition of droplet coalescence for the lowest $Re$ case (i.e., $Re = 5.2\times10^{3}$), it is reasonable to conclude that the similar behaviors of the droplet size observed here at high $Re$ have the same physical interpretations.
Moreover, similar results have been observed in additional experiments using a lower concentration of DBE ($12.5~\mu \text{L/L}$) (see appendix~\ref{appendix_D}).

\section{Conclusions}

In summary, we investigated the hydrodynamic behaviors of emulsions in a turbulent shear flow by varying the oil volume fraction from $0\%$ to $100\%$. 
Firstly, it is found that the average droplet diameter hardly changes with the oil volume fraction for O/W (or W/O) emulsions, while the ethanol-water droplets in W/O are $50\%$ larger than the oil droplets in O/W.
Secondly, the increasing trend of the effective viscosity versus the dispersed phase volume fraction is similar for both O/W and W/O emulsions, whereas the effective viscosity of O/W is found to be higher than that of W/O for the same Reynolds number, particularly for the case of high dispersed phase volume fractions. 
The asymmetric behaviors of the droplet size and the effective viscosity between O/W and W/O emulsions can be traced back to the presence of unavoidable surface-active contaminants, mainly from the wall, which probably preferentially dissolve in ethanol-water. 
In the presence of the contaminants, the coalescence of the oil droplets in O/W is suppressed when compared to the ethanol-water droplets with cleaner surface in W/O, yielding the smaller droplet size for O/W than that for its W/O counterpart.
Moreover, the higher effective viscosity of O/W than that of W/O can be connected to the smaller and non-deformable oil droplets due to the contaminants.
By introducing the oil-soluble surfactant DBE at a controlled concentration, we recover the symmetries of both the droplet size and the effective viscosity between O/W and W/O emulsions. This is consistent with the explanation of the mechanism responsible for the initial asymmetry.

Next, we discuss what sets the droplets size in turbulent emulsions.
Firstly, the normalized droplet size is found to be close to a $-6/5$ scaling dependence on the Reynolds number for the oil volume fraction of $1\%$ and of $99\%$, which is robust for both emulsions with and without surfactant.
Theoretically, the $-6/5$ scaling dependence can be obtained using either the K-H theory with the energy dissipation rate or the theory by Levich.
However, the Weber numbers being much less than 1 for the K-H theory indicates that the energy dissipation rate in the bulk flow is not enough to cause the breakup of such small droplets in this study.
According to the Weber number based on Levich’s theory, we conclude that the droplet fragmentation, which determines the droplet size, occurs within the boundary layer and is controlled by the dynamic pressure caused by the gradient of mean flow.

The present findings provide a better understanding of the hydrodynamic behaviors for both O/W emulsions and their W/O counterparts. The results on the effective viscosity open the possibility for active drag reduction during the oil recovery and transport through controlling the dispersed phase. Our finding of Levich’s droplet fragmentation mechanism also has some potential implications for the modulation of droplet size in chemical processing related to the dynamic emulsification. In future studies, more effects that may affect the effective viscosity and droplet size will be studied, aiming at attaining a complete understanding on the hydrodynamic behaviors of the turbulent emulsion at various conditions, and in particular near the phase inversion, where the phenomena are most striking.

\vspace{-4 mm}

\section*{Acknowledgements}
We acknowledge Mingbo Li, Sijia Lyu, and Stijn van Aartsen for their insightful suggestions and discussions. 
	We thank Oliver Masbernat for his help in selecting the surfactants.
	This work is financially supported by the Natural Science Foundation of China under grant nos.
	11988102 and 91852202, and the Tencent Foundation through the XPLORER PRIZE.

\vspace{-4 mm}
\section*{Declaration of interests}
The authors report no conflict of interest.

\appendix
\section{Liquids and surfactants}\label{appendix_A}

\begin{figure*}
	\centering
	\includegraphics[width = 1\textwidth]{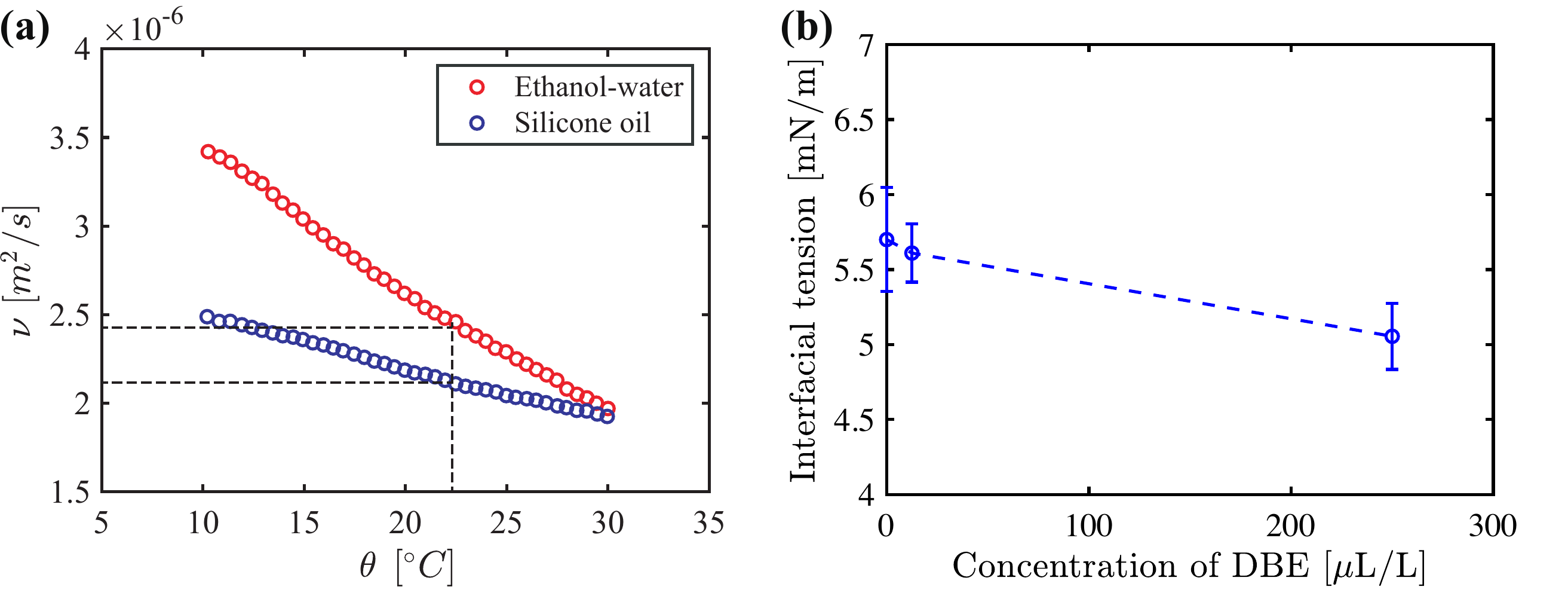}
	\caption{
		(a) The kinematic viscosity $\nu$ as a function of the temperature $\theta$. The red circles denote measured viscosity of ethanol-water, and the blue circles depict that of silicone oil. At the temperature for experiments $\theta=22^\circ$C, the viscosity of ethanol-water is $\nu_{w}=2.4\times10^{-6}\rm~m^{2}/s$, while that of silicone oil is $\nu_{o}=2.1\times10^{-6}\rm~m^{2}/s$. 
		(b) The interfacial tension as a function of the concentration of DBE in oil.
	}
	\label{fig5}
\end{figure*}

The silicone oil (KF-96L-2cSt), purchased from Shin-Etsu, has a viscosity of $\nu_{o}=2.1 \times10^{-6}\rm~m^{2}/s$ and a density of
$\rho_{o}=866\rm~kg/m^{3}$. The ethanol-water mixture ($\nu_{w}=2.4\times10^{-6}\rm~m^{2}/s$, $\rho_{w}=860\rm~kg/m^{3}$) is prepared with $25\%$ deionized water and $75\%$ ethanol in volume. Note that the viscosity of this mixture is very close to that of silicone oil. The viscosity values of both these two liquid phases are measured using a hybrid rheometer type of TA DHR-1 at a temperature of $\theta=22^\circ$C (see figure~\ref{fig5}(a)).

The density match of these two liquid phases eliminate the effect of centrifugal force on the liquid distribution. Furthermore, the dispersed droplets are expected to experience pressure fluctuations due to the strong turbulent liquid velocity fluctuations that develop in the current system. Here, we can compare the force induced by the velocity fluctuations to the centrifugal force by introducing a centrifugal Froude number~\citep{van2013importance};
\begin{equation}
Fr_{\text {cent }}(r)=\frac{\rho u^{\prime 2}_{\theta} / \left<D\right> }{\Delta \rho U_{\theta}^{2} / r},
\end{equation}
where $u^{\prime}_{\theta}$ denotes the standard deviation of the azimuthal liquid velocity fluctuations in the bulk, $U_{\theta}$ the mean azimuthal liquid velocity, and $r$ the radial position of the droplet to be considered. Based on the measurements in previous studies~\citep{grossmann2016high}, we take the estimation of $u^{\prime}_{\theta}\sim 0.01 \omega r_{i}$ and $U_{\theta}\sim 0.1\omega r_{i}$. Consequently, we find that $Fr_{\text {cent }}(r)$ is in the order of $10^2$, indicating that the centrifugal force is negligible compared to the force induced by the velocity fluctuations, which leads the droplets to get spread in the entire system.

A nonionic surfactants is used in experiments, which is a kind of dimethylsiloxane block copolymer (30-35\% Ethylene Oxide) purchased from Gelest. For convenience, we use its abbreviation (DBE) from the manufacturer. The density and molecular weight reported by the manufacturer are $970\rm~kg/m^{3}$ and around $10^{3}\rm~g/mol$, respectively. This surfactant is non-soluble in water but highly soluble in silicone oil. 
The interfacial tension between the two liquids (oil and ethanol-water) was measured using the pendent drop technique on a goniometer instrument (SCA20). Without using surfactant, the interfacial tension between oil and ethanol-water is $\gamma = 5.7~\text{mN/m}$. We performed measurements for emulsions containing DBE at various concentrations. As shown in figure~\ref{fig5}(b), the interfacial tension between the two liquids only slightly decreases with the increasing concentration of DBE.
Considering that the Levich prediction ($\left<D\right> = 2\sqrt{\gamma \nu /(25\rho u_{\ast}^{3})}$) gives a scaling between the droplet size and the interfacial tension as $\left<D\right> \sim \gamma^{1/2}$, we can estimate the droplet size reduction due to the interfacial tension reduction is around only $5\%$, which is much less than the $50\%$ jump of droplet size measured in experiments with adding 250~$\mu \text{L/L}$ DBE.
The effect of the interfacial tension reduction due to the DBE on the droplet size is unimportant in the current work.
Thus, it is reasonable to conclude that the inhibition of the droplet coalescence is the dominating factor in affecting the droplet size when using DBE in emulsions.

\section{The torque and the effective viscosity}\label{appendix_B}
\begin{figure*}
	\centering
	\includegraphics[width = 1\textwidth]{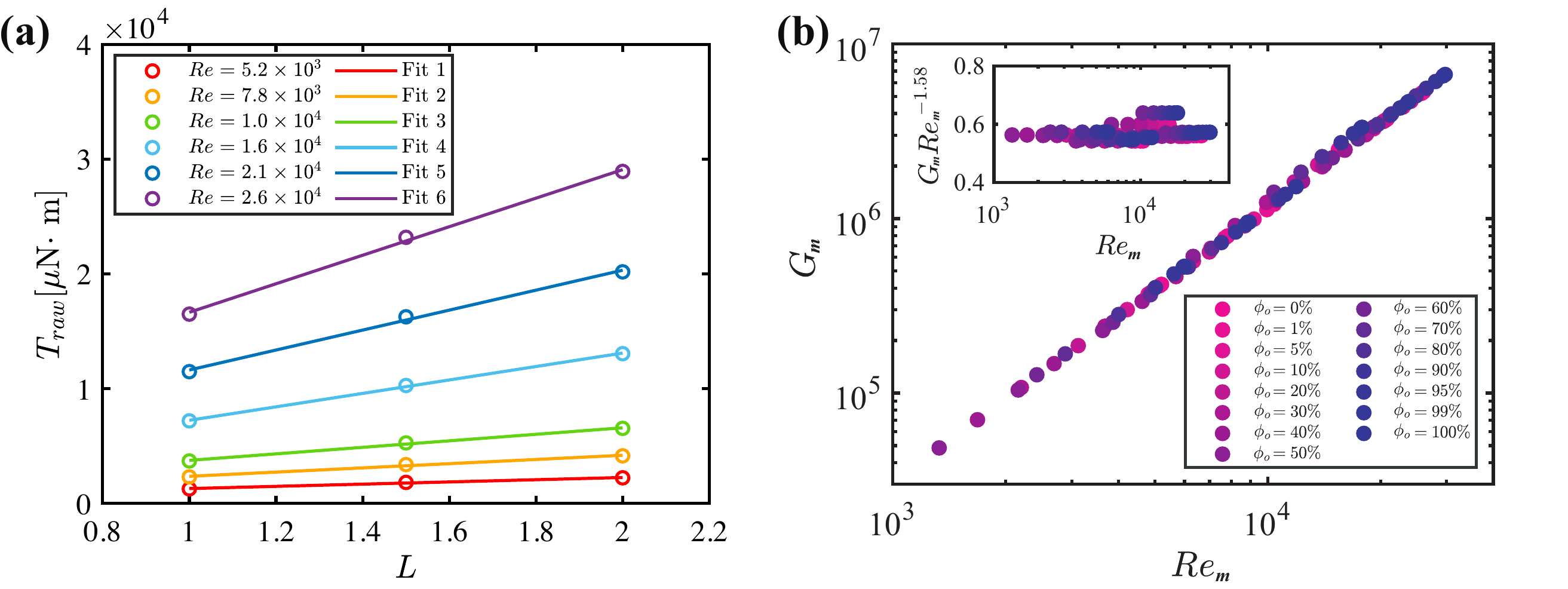}
	\caption{
		(a) The calibration of the torque measurements. The torque contribution of the end effect can be determined as the longitudinal intercept of the fitting line. Here, the oil volume fraction is $\phi_{o} = 0\%$.
		(b) The dependence between the modified dimensionless torque, $G_{m}$, and the modified Reynolds number, $Re_{m}$, by using the effective viscosity. All these sets of data at the various oil volume fractions collapse in a master curve, and the error is less than $1\%$. The inset shows the dimensionless torque compensated with $Re_{m}^{-1.58}$. Here, the results are from emulsions without using surfactant.
	}
	\label{fig6}
\end{figure*}

The torque is a response parameter of the emulsion system in the current study. The torque is directly measured by the rheometer through the shaft connected to the inner rotating cylinder with high accuracy up to $0.1\rm~nN\cdot m$. For each experiment, we set the angular velocity $\omega_{i}$ of the inner cylinder at a constant value. After the system reaches a statistically stable state, the measurement of the torque is performed. 

Note that the total torque ($T_{raw}$) directly measured by the torque sensor can be split up two parts: (1) the torque $T$ due to the sidewall of the inner cylinder (Taylor-Couette flow); (2) the torque $T_{end}$ related to the top and bottom end plates (von K\'am\'an flow). In this study, what we need is $T$, which can be determined by using a linearisation method~\citep{hu2017significant,greidanus2011drag}. We performed torque measurements in three TC devices with different heights of, $L$, $2L$, and $3L$. Since the contribution of the cylindrical sidewall increases linearly with the height of the cylinder~\citep{greidanus2011drag,hu2017significant}, we can obtain the $T_{end}$ as the longitudinal intercept of the linear fit (see figure~\ref{fig6}(a)). The ratio of the torque caused by the TC flow to the total torque can be given as $\beta = 1 - T_{end}/T_{raw}$, which is determined by performing experiments for two cases of single-phase flow (i.e., $\phi_{o} = 0\%$ and $\phi_{o} = 100\%$). Then, $\beta$ obtained can be applied to the flow with internal dispersed phase (i.e., emulsions). Consequently, the value of the torque caused by the TC emulsion flow can be calculated for various oil volume fractions.

For a TC turbulent emulsion, the control parameter can be defined using the modified Reynolds number,

\begin{equation}
	Re_{m} = \omega_{i}r_id/\nu_{eff}, 
\end{equation}
where $\nu_{eff}$ is the effective viscosity of the emulsion.
The response parameter is the modified dimensionless torque given by: 

\begin{equation}
	G_{m}=T/2\pi L\rho\nu_{eff}^{2}, 
\end{equation}
where $T$ denotes the torque that is required to maintain the inner cylinder rotating at a constant angular velocity $\omega_i$.

Firstly, we calculate the $Re_{m}$ and $G_{m}$ at various angular velocities $\omega_i$ for pure ethanol-water mixture ($\phi_{o}=0\%$) with a known viscosity. When we plot these data in a $G_{m}-Re_{m}$ plot, we find a scaling law as $G_{m}\sim Re_{m}^{1.58}$ (see figure~\ref{fig6}(b)). Further, we can write this relation as $G_{m}=KRe_{m}^{1.58}$, where $K$ denotes a constant prefactor. If we insert the definitions of $G_{m}$ and $Re_{m}$ to this dependence, we obtain a dependence of torque $T$ and viscosity $\nu_{eff}$ as

\begin{equation}
	T=AK\nu_{eff}^{0.42},
\end{equation}
where $A$ equals to $2\pi L\rho(\omega_{i}r_id)^{1.58}$.
This relation is expected to be valid for emulsion systems with various oil volume fractions and Reynolds numbers as well, which is supported by previous studies~\citep{ravelet2007experimental,bakhuis2021catastrophic}.  The torque and the effective viscosity of the emulsion system can be denoted as $T$ and $\nu_{eff}$ for a constant angular velocity $\omega_i$ at the oil volume fraction of $\phi_{o}$. For the pure ethanol-water mixture ($\phi_{o}=0\%$) system at the same angular velocity, we obtain the measured torque value $T_{w}$ and the viscosity $\nu_{w}$. Both these two systems follow the relation given above. Since the angular velocities of these two systems are the same, the prefactor $A$ is the same too. Then, we can derive the following relation:

\begin{equation}
	\frac{\nu_{eff}}{\nu_{w}}=\Big(\frac{T}{T_{w}}\Big)^{2.38}.
\end{equation}
The effective viscosity of emulsion systems $\nu_{eff}$ can be obtained based on this relation.
By using the effective viscosity obtained for each cases, we calculate $G_{m}$ and $Re_{m}$ for various volume fractions and angular velocities. When we plot together all data in a $G_{m}$-$Re_{m}$ plot, we find that all data sets of $G_{m}$ versus $Re_{m}$ collapse in a master curve, for various oil fractions (see figure~\ref{fig6}(b)). 

\section{The dependence of the droplet size on the Reynolds number}\label{appendix_C}
In a TC turbulent flow, the boundary layer thickness can be estimated as:

\begin{equation}
	\lambda_{u} = d/(2Nu_{\omega}).
\end{equation}
Here, we use another typical response parameter in a TC turbulent flow, the angular velocity Nusselt number:

\begin{equation}
	Nu_{\omega} = \frac{T}{2\pi L\rho J_{\omega}^{lam}},
\end{equation}
where $J_{\omega}^{lam} = 2\nu r_{i}^2 r_{o}^2\omega_{i}/(r_{o}^2 - r_{i}^2)$ is the angular velocity transport for the laminar TC flow.
Based on this, the boundary layer thickness, $\lambda_{u}$, is found to be around 5 times larger than droplet size, $\left<D\right>$, for various Reynolds numbers, which supports the Levich theory in the main paper.

It is found that the droplet breakup is most often observed close to the area where the most intense stress participates in the deformation~\citep{hesketh1991experimental,afshar2013liquid}. 
Thus, the droplet size is expected to be dominated by the boundary layer region close to the wall, where the energy dissipation rate ($\varepsilon_{bl}$) is largest~\citep{ezeta2018turbulence}. 
The $\varepsilon_{bl}$ can be estimated as $\varepsilon_{bl} = u_{\ast}^{3}/\lambda_{u}$, where we use the shear velocity: $u_{\ast} = \sqrt{\tau_w/\rho} = \sqrt{T/(2\pi\rho r_{i}^{2}L)}$ and the boundary layer thickness: $\lambda_{u} = d/(2Nu_{\omega})$ in a TC turbulent flow~\citep{eckhardt2007torque}. 
Thus, the normalized energy dissipation rate can be estimated as $\varepsilon_{bl}/(\nu^3 d^{-4}) \sim G^{3/2}Nu_{\omega}$.
Since the volume fraction of the dispersed phase is very low here ($\phi_{o} = 1\%$ or $\phi_{o} = 99\%$), the viscosity of the emulsion is approximately equal to that of the continuous phase, giving $Re_{m} \approx Re$ and $G_{m} \approx G$.
Using the effective scaling of $G = T/2\pi L\rho\nu^{2} \sim Re^{1.58}$ obtained in the current system (see figure~\ref{fig6}(b)), we can get $Nu_{\omega} \sim Re^{0.58}$.
Consequently, the energy dissipation rate is found to scale as $\varepsilon_{bl}/(\nu^3 d^{-4}) \sim Re^{2.95}$. 
Inserting this expression of $\varepsilon_{bl}$ into the K-H prediction (i.e., $\left<D\right> = C(\rho/\gamma)^{-3/5}\varepsilon_{bl}^{-2/5}$), one obtains $\left<D\right>/d \sim Re^{-1.18}$, which is also in good agreement with the experimental data. The scaling exponent of $-1.18$, which is close to the $-6/5$ given by the K-H theory using dissipation rate in the bulk, suggests that the energy dissipation at the boundary layer ($\varepsilon_{bl}$) is just proportional to the local energy dissipation in the bulk of the system ($\varepsilon_{b}$). Note that this is similar to what observed for the case of the liquid-liquid dispersion in an agitated vessel~\citep{wichterle1995drop}. 

However, the Weber numbers calculated using $\varepsilon_{bl}$ are about $0.08$, which is an order magnitude smaller than the critical value ($\mathcal{O}(1)$). This indicates that the energy dissipation rate $\varepsilon_{bl}$ near the wall is not large enough to cause the breakup of such small droplets. These results again suggest that the K-H theory is not appropriate for modeling the droplet size in the present system.

\section{Additional experiments}\label{appendix_D}

\begin{figure*}
	\centering
	\includegraphics[width = 1\textwidth]{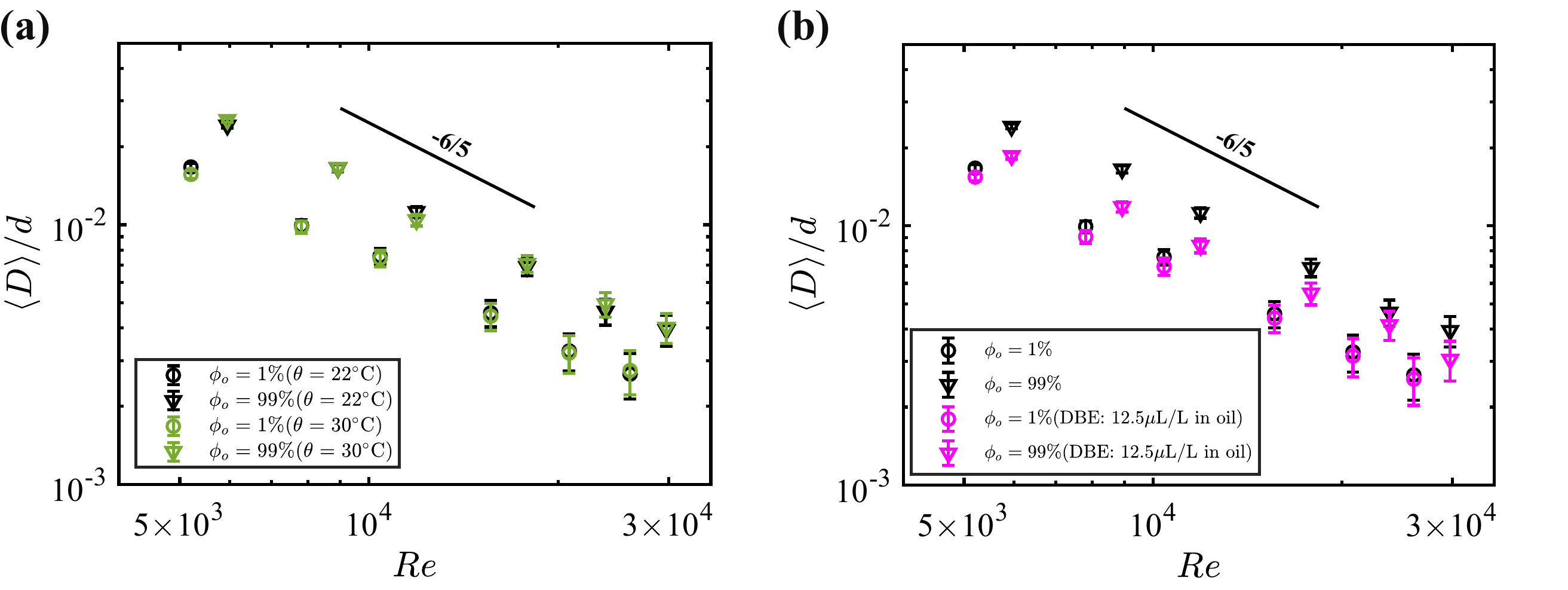}
	\caption{
	(a) The droplet size for various Reynolds numbers at $\theta=22^\circ$C and at $\theta=30^\circ$C.
	(b) The average droplet diameter normalized by the gap width as a function of the Reynolds number for oil volume fractions of $\phi_{o}=1\%$ and of $99\%$. The black and magenta symbols denote droplet sizes in emulsions using no surfactant and using $12.5~\mu \text{L/L}$ DBE, respectively.
		The black lines denote the scaling prediction of the Kolmogorov-Hinze theory. The asymmetry of the droplet size in O/W ($\phi_{o}=1\%$) and W/O ($\phi_{o}=99\%$) emulsions is partially removed using $12.5~\mu \text{L/L}$ DBE.
	}
	\label{fig7}
\end{figure*}
One may think that the difference of the viscosity between oil ($\nu_{o}=2.1\times10^{-6}\rm~m^{2}/s$) and ethanol-water ($\nu_{w}=2.4\times10^{-6}\rm~m^{2}/s$) could be the source of the asymmetry of the droplet size. By adjusting the temperature of the emulsion from $\theta=22^\circ$C to $\theta=30^\circ$C, we eliminate the viscosity difference between the two liquids (see figure~\ref{fig5}(a)) and measure the droplet size at oil volume fraction of $\phi_{o}=1\%$ and of $\phi_{o}=99\%$. As shown in figure~\ref{fig7}(a), the droplet sizes only have slight change when compared to the results obtained at $\theta=22^\circ$C. Clearly, the ethanol-water droplets in W/O are larger than the oil droplets in O/W as well, for all Reynolds numbers.
Thus, this small viscosity difference between the two liquids used in experiments cannot account for the obvious asymmetry of the droplet size.

Here, we provide experimental results for emulsions using $12.5~\mu \text{L/L}$ DBE. As shown in figure~\ref{fig7}(b), the -6/5 scaling dependence of the droplet size on the Reynolds number remains, suggesting the robustness of the scaling law for turbulent emulsions containing $12.5~\mu \text{L/L}$ DBE.
For the case of $\phi_{o}=1\%$, it is found that the droplet size only slightly decreases due to DBE. However, the droplet size at $\phi_{o}=99\%$ shows an obvious reduction, yielding the partial elimination of the asymmetry of droplet size, for various Reynolds numbers.

\vspace{-2 mm}
\bibliographystyle{jfm} 

\end{document}